\documentclass{article}
\usepackage{spconf,amsmath,graphicx}
\usepackage{txfonts}
\usepackage{enumitem}
\usepackage{color,xcolor}
\usepackage{colortbl}
\usepackage{threeparttable}
\usepackage{subfigure}
\usepackage[colorlinks=true]{hyperref}
\usepackage{array}
\newcommand{\PreserveBackslash}[1]{\let\temp=\\#1\let\\=\temp}
\newcolumntype{C}[1]{>{\PreserveBackslash\centering}p{#1}}
\newcolumntype{R}[1]{>{\PreserveBackslash\raggedleft}p{#1}}
\newcolumntype{L}[1]{>{\PreserveBackslash\raggedright}p{#1}}

\title{Deep Generative Fixed-filter Active Noise Control}

\name{Zhengding Luo, Dongyuan Shi, Xiaoyi Shen, Junwei Ji, Woon-Seng Gan\thanks{The code is available at \href{https://github.com/Luo-Zhengding/GFANC}{https://github.com/Luo-Zhengding/GFANC}}}
\address{School of Electrical \& Electronic Engineering, Nanyang Technological University, Singapore.\\
Emails: LUOZ0021@e.ntu.edu.sg; dongyuan.shi@ntu.edu.sg; ewsgan@ntu.edu.sg}

\begin{document}
%\ninept % 整体压缩行距
\maketitle

\begin{abstract}
Due to the slow convergence and poor tracking ability, conventional LMS-based adaptive algorithms are less capable of handling dynamic noises. Selective fixed-filter active noise control (SFANC) can significantly reduce response time by selecting appropriate pre-trained control filters for different noises. Nonetheless, the limited number of pre-trained control filters may affect noise reduction performance, especially when the incoming noise differs much from the initial noises during pre-training. Therefore, a generative fixed-filter active noise control (GFANC) method is proposed in this paper to overcome the limitation. Based on deep learning and a perfect-reconstruction filter bank, the GFANC method only requires a few prior data (one pre-trained broadband control filter) to automatically generate suitable control filters for various noises. The efficacy of the GFANC method is demonstrated by numerical simulations on real-recorded noises.
\end{abstract}

\vspace*{-0.3cm}

\section{Introduction}\vspace*{-0.2cm}
Active noise control (ANC) is widely regarded as an effective noise cancellation technique that produces an "anti-noise" from a secondary source to suppress the unwanted noise \cite{1,liebich2018signal,2,qiu2019introduction}. Based on the principle of superposition of acoustic signals, the anti-noise is of equal amplitude and opposite phase to the disturbance \cite{3}. ANC is particularly effective at cancelling low-frequency noises, where passive methods are typically ineffective, prohibitively expensive or cumbersome \cite{4}. Therefore, ANC technologies have been integrated into a wide range of commercial products including headphones, mobile phones, automobiles, etc.

In adaptive ANC systems, LMS-based algorithms such as filtered-X least mean square (FxLMS) are typically employed \cite{5,6}. Nevertheless, they are less capable of handling rapidly varying or non-stationary noises due to the inherent slow convergence and poor tracking ability. Moreover, their slow response speed and risk of divergence also impact customers' perceptions of the ANC effect \cite{7}. To improve the response speed and robustness, many audio products adopt the fixed-filter ANC methods \cite{8}. However, the fixed control filter is only suitable for a specific noise type, resulting in degraded noise reduction performance for other types of noises \cite{27}.\vspace*{-0.05cm}

To select different pre-trained control filters given different noise types, a selective fixed-filter active noise control (SFANC) method based on frequency band matching was proposed in \cite{9}. However, the critical parameters of the method can only be determined by trial and error, limiting its practical applications. Thus, convolutional neural networks (CNNs) are integrated into the SFANC approach \cite{10}, where all the parameters are automatically learned from the noise dataset. Nevertheless, the number of pre-trained control filters in the SFANC approach is limited, resulting in poor noise control performance for some noises, particularly when they are quite distinct from the filter-training noises.\vspace*{-0.05cm}

To overcome the limitation of SFANC and generate suitable control filters for various noises, a deep generative fixed-filter active noise control (GFANC) method is proposed in this paper. Firstly, we apply the theory of filter perfect-reconstruction \cite{24} to decompose a pre-trained broadband control filter into multiple sub control filters. On a co-processor, a lightweight one-dimensional CNN (1D CNN) automatically provides the combination weights of sub control filters given the primary noise. After that, a new control filter is generated by the weighted sum of sub control filters and then used for noise cancellation in the real-time controller. For training the 1D CNN, a novel adaptive labelling mechanism is developed to label the training noises without extra human efforts.\vspace*{-0.05cm}

The operation on the co-processor (e.g., a mobile phone) can alleviate the computational load of the real-time noise controller. Also, the GFANC approach only requires a single pre-trained control filter, which significantly reduces the ANC's preparation efforts and improves its practicality. Additionally, different from conventional adaptive ANC algorithms, the absence of a feedback error signal \cite{19} in the GFANC method can minimize the risk of divergence. The numerical simulations on real-recorded noises show that the GFANC method can achieve an excellent balance between noise reduction performance and response time.\vspace*{-0.4cm}

\section{The Proposed GFANC Approach}
\begin{figure}[tp]
\centering
\centerline{\includegraphics[width=\linewidth]{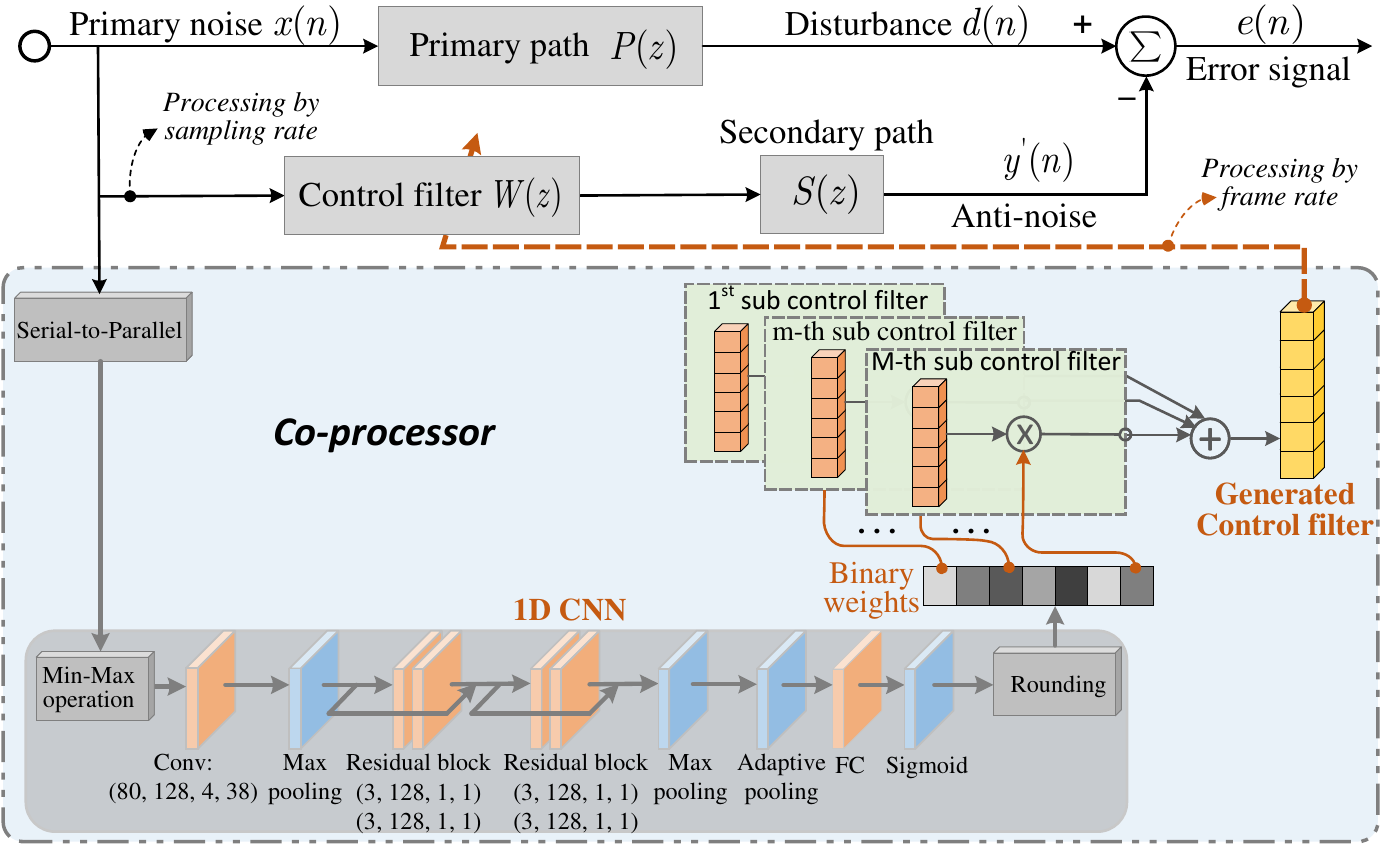}}\vspace*{-0.3cm}
\caption{Block diagram of the proposed GFANC approach. Convolution layer: (kernel size, channels, stride, padding).}
\label{Fig 1}\vspace*{-0.5cm}
\end{figure}

\vspace*{-0.2cm}
Fig.~\ref{Fig 1} illustrates the GFANC approach's overall architecture, where the co-processor uses the 1D CNN to provide a binary weight vector given each frame's noise data. Through the inner product of this binary weight vector and fixed sub control filters, a new control filter is generated and then utilized for real-time noise control. The weights used are binary weights to reduce the load of computation and memory space. Especially, the co-processor operates at the frame rate, while the real-time controller operates at the sampling rate.\vspace*{-0.5cm}

\subsection{Construction of Sub Control Filters}\vspace*{-0.1cm}
The construction of sub control filters is an essential step of the proposed method. We used a practical filter decomposition technique based on the theory of filter perfect-reconstruction \cite{20} to obtain sub control filters. Firstly, we used the target ANC system to cancel a broadband primary noise $\mathbf{x}(n)$ containing the frequency components of interest. Its optimal control filter determined by the FxLMS algorithm \cite{22} is our pre-trained broadband control filter.

It is assumed that the pre-trained broadband control filter has $N$ taps, which is represented as $\mathbf{c}=[c(0),\cdots,c(N-1)]^\mathrm{T}$. $\mathbf{c}$ will be decomposed into a perfect-reconstruction filter bank as outlined below. Through the discrete Fourier transform (DFT), its frequency spectrum is derived as
%-----------------------------------------------------------------------
\begin{small}
\begin{equation}\label{equation 1}
\setlength{\abovedisplayskip}{2pt}
\setlength{\belowdisplayskip}{2pt}
\mathbf{C}=\mathbf{F}_{N}\mathbf{c}=[C(0),\cdots,C(k),\cdots,C(N-1)]^\mathrm{T},
\end{equation}
\end{small}
where $\mathbf{F}_{N}$ denotes the DFT matrix. According to the conjugate symmetry of the real signal in frequency domain, the control filter $\mathbf{C}$ is divided into $M$ sub control filters as
\begin{small}
\begin{equation}\label{equation 2}
\setlength{\abovedisplayskip}{2pt}
\setlength{\belowdisplayskip}{2pt}
\mathbf{C}=\sum^{M}_{m=1}\mathbf{C}_m,
\end{equation}
\end{small}
in which the frequency spectrum of $m$-th sub control filter can be represented as
\begin{small}
\begin{equation}\label{equation 3}
\setlength{\abovedisplayskip}{2pt}
\setlength{\belowdisplayskip}{2pt}
\mathbf{C}_m=[C_m(0),\cdots,C_m(k),\cdots,C_m(N-1)]^\mathrm{T}~~~m\in[1, M].
\end{equation}
\end{small}
When $m\ne M$ the elements in \eqref{equation 3} are obtained from
\begin{small}
\begin{equation}
\setlength{\abovedisplayskip}{2pt}
\setlength{\belowdisplayskip}{2pt}
C_m(k)=\begin{cases}C(k)&k\in[(m-1)I+1, mI]\\&\cup[N-mI,N-1-(m-1)I]\\
0&\text{others;}\end{cases}
\end{equation}
\end{small}
When $m=M$, the sub control filter’s frequency spectrum is
\begin{small}\begin{equation}
\setlength{\abovedisplayskip}{2pt}
\setlength{\belowdisplayskip}{2pt}
C_M(k)=\begin{cases}C(k)&k\in[(M-1)I+1, N-1-(M-1)I]\\0&\text{others,}\end{cases}
\end{equation}
\end{small}
where the bandwidth of sub control filter is $I=\lfloor\frac{N}{2M}\rfloor$.

The time-domain representation of the $m$-th sub control filter is computed by
\begin{small}
\begin{equation}\label{equation 6}
\setlength{\abovedisplayskip}{2pt}
\setlength{\belowdisplayskip}{2pt}
\mathbf{c}_m=\mathbf{F}^{-1}_{N}\mathbf{C}_m,
\end{equation}
\end{small}
where $\mathbf{F}^{-1}_{N}$ stands for the inverse DFT matrix. 

Furthermore, by combing  \eqref{equation 1}, \eqref{equation 2}, and \eqref{equation 6}, the control signal $y(n)$ can be written as
\begin{small}
\begin{equation}
\setlength{\abovedisplayskip}{2pt}
\setlength{\belowdisplayskip}{2pt}
\begin{split}y(n)&=\mathbf{x}^\mathrm{T}(n)\mathbf{c}=\mathbf{x}^\mathrm{T}(n)\mathbf{F}^{-1}_{N}\mathbf{F}_{N}\mathbf{c}=\mathbf{x}^\mathrm{T}(n)\mathbf{F}^{-1}_{N}\mathbf{C}\\&=\mathbf{x}^\mathrm{T}(n)\mathbf{F}^{-1}_{N}\sum^{M}_{m=1}\mathbf{C}_m=\mathbf{x}^\mathrm{T}(n)\sum^{M}_{m=1}\mathbf{c}_m.\end{split}
\end{equation}
\end{small}

It indicates that the control filter $\mathbf{c}$ can be perfectly reconstructed in time domain by its all sub-band filters $\mathbf{c}_m$. Therefore, by decomposing the pre-trained broadband control filter based on the theory of filter perfect-reconstruction, multiple fixed sub control filters can be obtained.\vspace*{-0.4cm}

\begin{figure}[tp]
\centering
\centerline{\includegraphics[height=3.5cm]{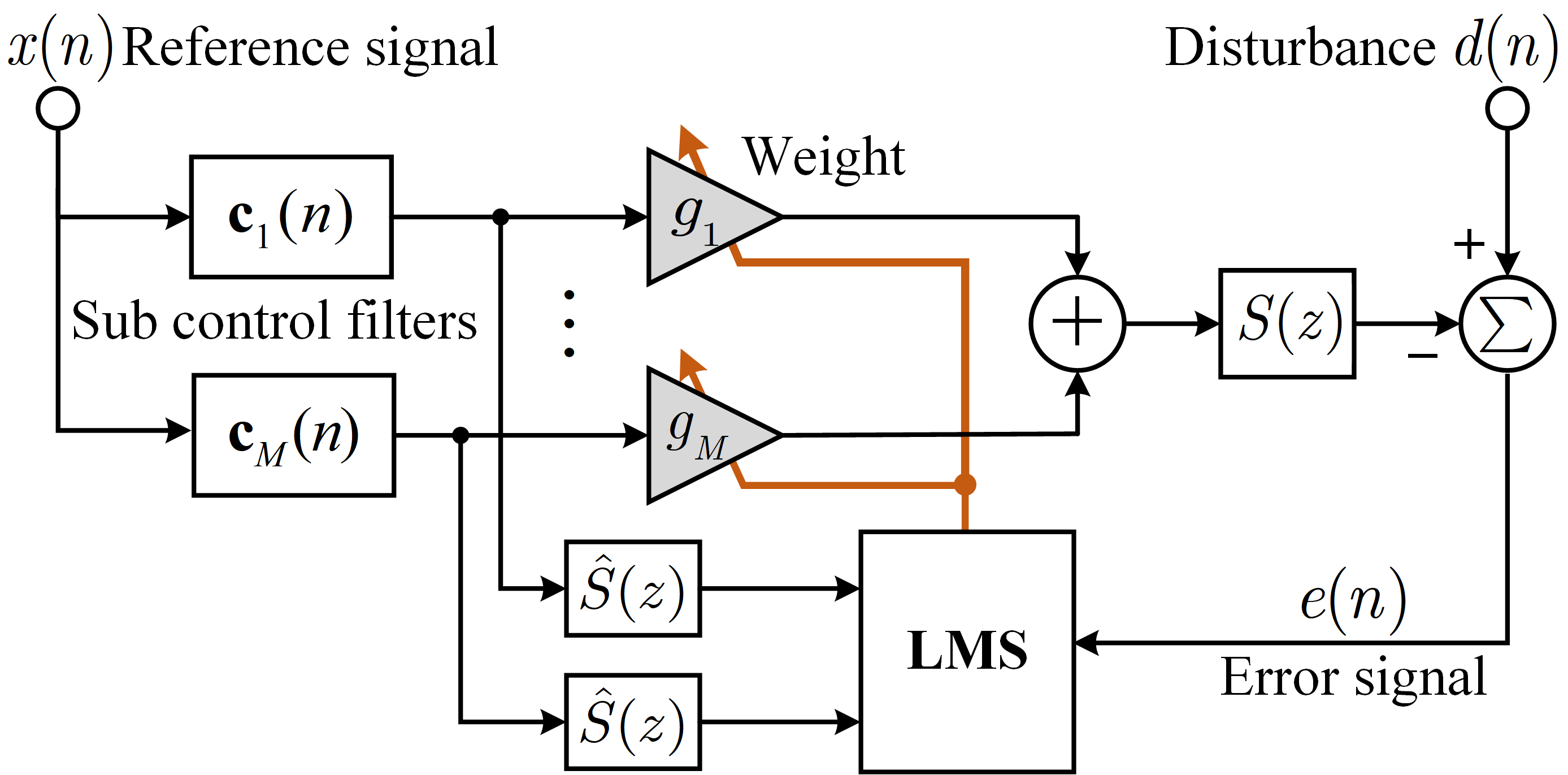}}\vspace*{-0.3cm}
\caption{Adaptive labelling mechanism to get the optimal combination weights of sub control filters given the training noise.}
\label{Fig 2}\vspace*{-0.4cm}
\end{figure}

\subsection{Adaptive Labelling Mechanism}
The input and output of the 1D CNN are a frame of 1-second noise and its corresponding binary weight vector, respectively. For automatically labelling the noise dataset, we proposed an adaptive labelling mechanism as shown in Fig.~\ref{Fig 2} to reduce the required time and effort. Formally, the training noise is input as the reference signal $\mathbf{x}(n)$.

Hence, the output signal of the $m$-th fixed sub control filter is computed to be
\begin{small}
\begin{equation}
\setlength{\abovedisplayskip}{2pt}
\setlength{\belowdisplayskip}{2pt}
    y_{m}(n)=\mathbf{x}^\mathrm{T}(n) \mathbf{c}_{m}~~~m\in[1, M].
\end{equation}
\end{small}

The control signal $y_g(n)$ is subsequently derived from the weighted sum of all sub control filter outputs:
\begin{small}
\begin{equation}
\setlength{\abovedisplayskip}{2pt}
\setlength{\belowdisplayskip}{2pt}
    y_g(n)=\mathbf{y}^{\mathrm{T}}(n) \mathbf{g}(n),
\end{equation}
\end{small}
where \begin{small}$\mathbf{y}(n)=\left[y_{1}(n), \cdots, y_{m}(n), \cdots, y_{M}(n)\right]^{\mathrm{T}}$\end{small}, and the combination weight vector is \begin{small}$\mathbf{g}(n)=\left[g_{1}(n), \cdots, g_{m}(n), \cdots, g_{M}(n) \right]^{\mathrm{T}}$\end{small}.

The residual error signal is calculated as
\begin{small}
\begin{equation}\label{Residual_error}
\setlength{\abovedisplayskip}{2pt}
\setlength{\belowdisplayskip}{2pt}
    e(n)=d(n)-y_g(n)*s(n),
\end{equation}
\end{small}
where $d(n)$ and $s(n)$ represent the disturbance collected by the error microphone and the impulse response of secondary path $S(z)$, respectively. $*$ represents the linear convolution.

To minimize the square of \eqref{Residual_error} through LMS algorithm \cite{25}, the updating formula of the weights is derived as
\begin{small}
\begin{equation}\label{Aaptive_label}
\setlength{\abovedisplayskip}{2pt}
\setlength{\belowdisplayskip}{2pt}
\begin{aligned}
    \mathbf{g}(n+1)&=\mathbf{g}(n)+\mu \mathbf{y}^\prime(n)e(n),\\
    \mathbf{y}^\prime(n)&=\mathbf{y}(n)*s(n),
\end{aligned}
\end{equation}
\end{small}
where $\mu$ stands for the updating step size.

Once \eqref{Aaptive_label} converges, the binary weight vector $\mathbf{T}$ is gained by rounding the optimal weight vector:
\begin{small}
\begin{equation}
\setlength{\abovedisplayskip}{2pt}
\setlength{\belowdisplayskip}{2pt}
\begin{aligned}
t_m &=
\begin{cases}
1 & g^o_m \ge 0.5,\\
0 & g^o_m < 0.5,
\end{cases} \\
\mathbf{T} &=\left[t_1, \cdots, t_m,\cdots, t_M\right]^\mathrm{T},
\end{aligned}
\end{equation}
\end{small}
where $g^o_m$ means the optimal combination weight of $m$-th sub control filter. Finally, the automatically obtained binary weight vector $\mathbf{T}$ is seen as the label of the training noise.\vspace*{-0.4cm}

\subsection{The Proposed 1D CNN}
The architecture of the proposed 1D CNN used for getting the combination weights is shown in Fig.~\ref{Fig 1}. In the pre-processing stage, there is a min-max operation defined as
\begin{small}
\begin{equation}
\setlength{\abovedisplayskip}{2pt}
\setlength{\belowdisplayskip}{2pt}
\hat{x}(n)=\frac{x(n)}{\max [\mathbf{x}(n)]-\min [\mathbf{x}(n)]},
\end{equation}
\end{small}
where $\max [\cdot]$ and $\min [\cdot]$ mean obtaining the maximum and minimum value of the input vector $\mathbf{x}(n)$, respectively. The operation rescales the input range into $(-1,1)$ and retains its negative part containing phase information, which is important for ANC applications \cite{21}.

Motivated by the work \cite{12}, the 1D CNN is designed as a lightweight network including two residual blocks. Every residual block comprises two convolutional layers, batch normalization and ReLU non-linearity. Also, a shortcut connection is utilized in each residual block, as the residual architecture has been shown to be easily optimized \cite{13}. Additionally, the network uses a broad receptive field (RF) in the first convolutional layer and narrow RFs in the rest convolutional layers to fully exploit both global and local information.

In this task, obtaining the binary weight vector belongs to a multi-label classification problem \cite{14}. Thus, Sigmoid function is used as the output layer of the 1D CNN. Also, Binary Cross Entropy (BCE) loss is applied to train the model.\vspace*{-0.5cm}

\begin{figure}[tp]
\subfigure[A broadband control filter]{
\centering
\includegraphics[height=3.65cm]{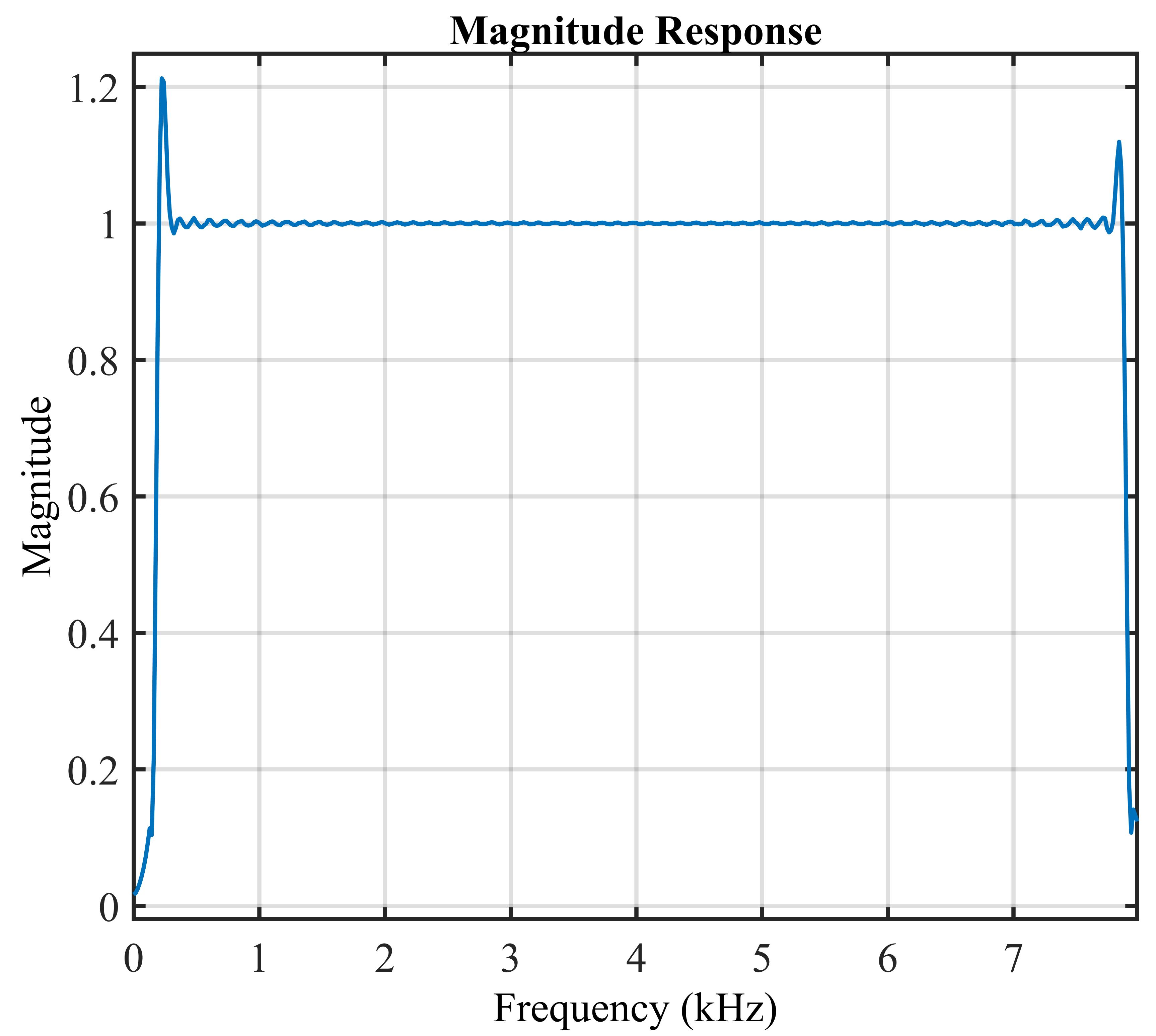}
}
\subfigure[Sub control filters]{
\centering
\includegraphics[height=3.65cm]{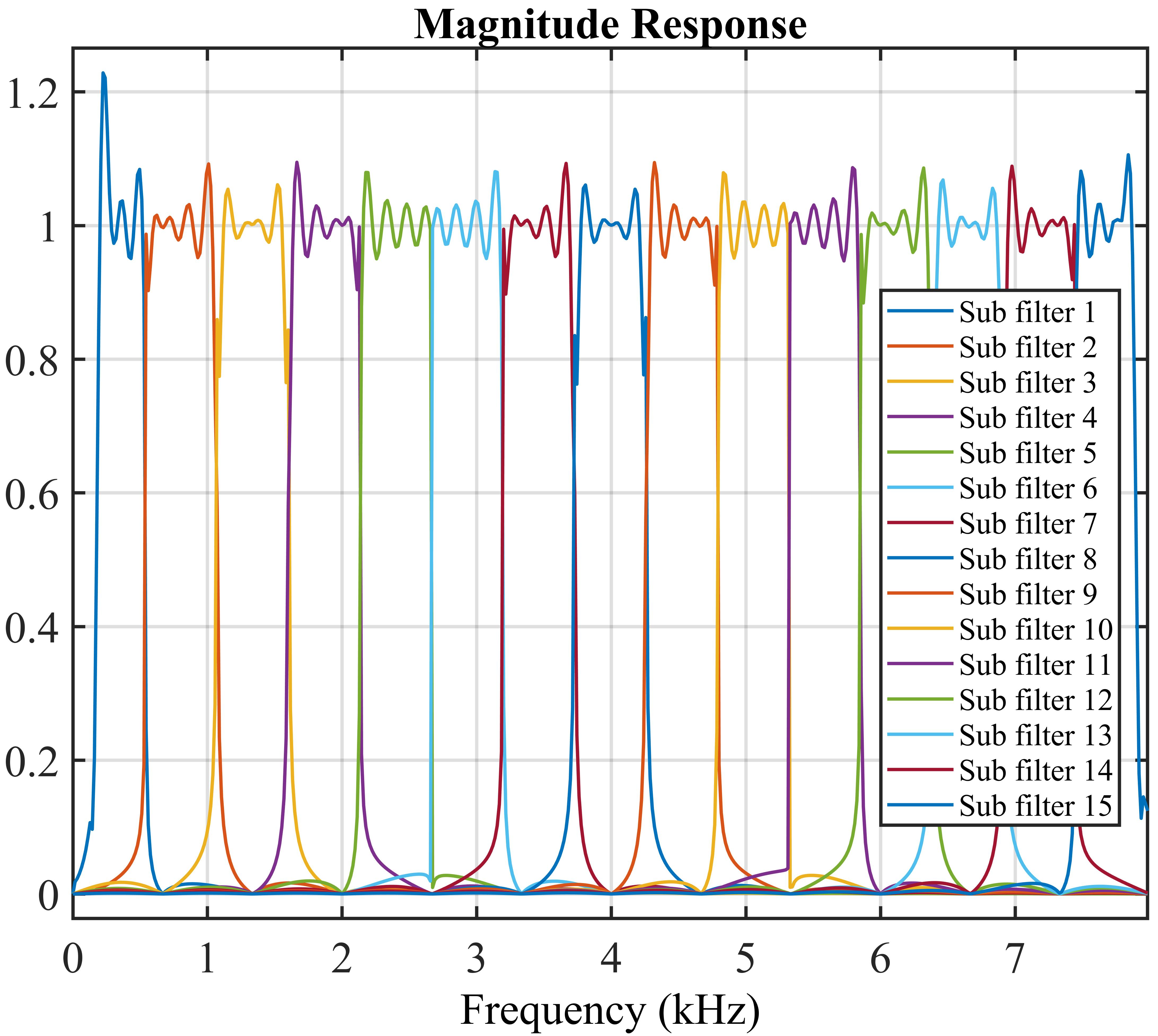}
}\vspace*{-0.3cm}
\caption{The sub control filters used in the GFANC approach.}
\label{Fig 3}\vspace*{-0.4cm}
\end{figure}

\subsection{Generation of Control Filter}\vspace*{-0.1cm}
After training, the 1D CNN can output the binary weight vector depending on the primary noise. Then, a new control filter is formed by the weighted sum of sub control filters. Formally, the generated control filter $\mathbf{\hat{w}}$ is obtained from
\begin{small}
\begin{equation}
\setlength{\abovedisplayskip}{2pt}
\setlength{\belowdisplayskip}{2pt}
    \mathbf{\hat{w}}=\sum_{m=1}^{M}t_m\cdot\mathbf{c}_m~~~m\in[1, M].
\end{equation}
\end{small}
The generated control filter would be used by the real-time processor for noise cancellation:
\begin{small}
\begin{equation}
\setlength{\abovedisplayskip}{2pt}
\setlength{\belowdisplayskip}{2pt}
    e(n)= d(n)-\mathbf{x}^{\mathrm{T}}(n)\mathbf{\hat{w}}\ast s(n).
\end{equation}
\end{small}

Overall, given the incoming noise, a new control filter can be generated by adaptively combining fixed sub control filters. Instead of selecting from a limited set of control filters, GFANC can generate a more suitable control filter for each type of noise compared to SFANC. Moreover, unlike adaptive ANC algorithms, GFANC does not use the error signal to update the control filter, reducing the risk of divergence.\vspace*{-0.5cm}

\begin{footnotesize}
\begin{table}[!t]
\caption{Performance comparison of different networks.}
\begin{center}
\begin{tabular}{|L{2.7cm}C{2.6cm}C{1.8cm}|} %调整表格宽度并居中
\hline
\textbf{Network} & \textbf{Testing Accuracy} & \textbf{\#Parameters} \\
\hline
\rowcolor{gray!20}
\multicolumn{3}{|c|}{1D Convolutional Neural Networks} \\
\hline
Proposed Network & \textbf{97.20$\%$} & \textbf{0.21M} \\
M3 Network & 96.64$\%$ & 0.22M \\
M5 Network & 96.70$\%$ & 0.56M \\
M11 Network & 96.65$\%$ & 1.79M \\
M18 Network & 94.35$\%$ & 3.69M \\
M34-res Network & 96.74$\%$ & 3.99M \\
\hline
\rowcolor{gray!20}
\multicolumn{3}{|c|}{2D Convolutional Neural Networks} \\
\hline
ShuffleNet v2 & 96.40$\%$ & 0.25M \\
Mobilenet v2 & 96.05$\%$ & 2.89M \\
Attention Network & 95.80$\%$ & 4.95M \\
\hline
\end{tabular}
\label{Table 1}
\end{center}\vspace*{-0.8cm}
\end{table}
\end{footnotesize}

\section{Simulation Results}\vspace*{-0.2cm}
In the simulations, the pre-trained broadband control filter is divided into 15 sub control filters as depicted in Fig.~\ref{Fig 3}. The length of the control filter and the sampling rate are set to $1,024$ taps and $16$kHz, respectively. The primary and secondary paths used are bandpass filters with a frequency range of $20$Hz-$7,980$Hz. For training the network, a synthetic noise dataset divided into 3 subsets: $80,000$ noise tracks for training, $2,000$ noise tracks for validation, and $2,000$ noise tracks for testing is used. Each noise track has a 1-second duration.\vspace*{-0.5cm}

\subsection{Comparison of Different Networks}\vspace*{-0.1cm}
The proposed 1D CNN is compared to several 1D networks \cite{12}: M3, M5, M11, M18, and M34-res. Also, some light 2D networks including ShuffleNet v2 \cite{15}, Mobilenet v2 \cite{16} and Attention Network \cite{17} are used as baselines. The performance of these networks in the GFANC approach are summarized in Table \ref{Table 1}, where the testing accuracy means the accuracy of predicting binary weights on the test dataset.

On the test dataset, the proposed 1D network achieves the highest prediction accuracy of $97.20\%$, indicating that it can effectively extract discriminative features from raw waveforms to provide suitable binary weights given different noises. Meanwhile, the proposed 1D network is lightweight, with only $0.21M$ parameters, making it suitable to be implemented on co-processors such as mobile phones. Furthermore, compared to using frequency-domain data in 2D networks, using time-domain waveform directly in 1D networks is more convenient and practical \cite{23}.

\begin{figure}[!t]
\centering
\subfigure{
\includegraphics[width=0.455\linewidth]{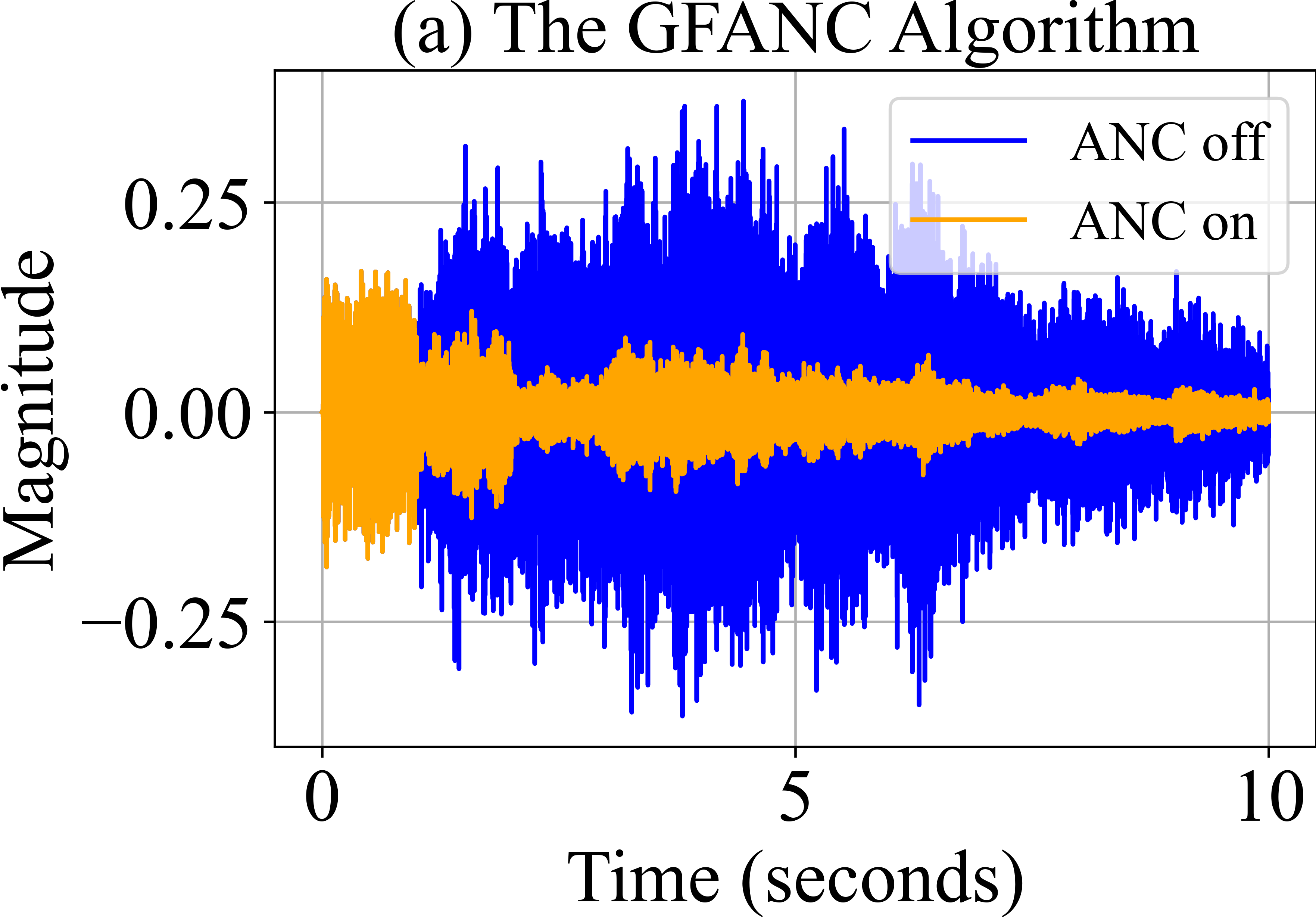}
}
\subfigure{
\includegraphics[width=0.455\linewidth]{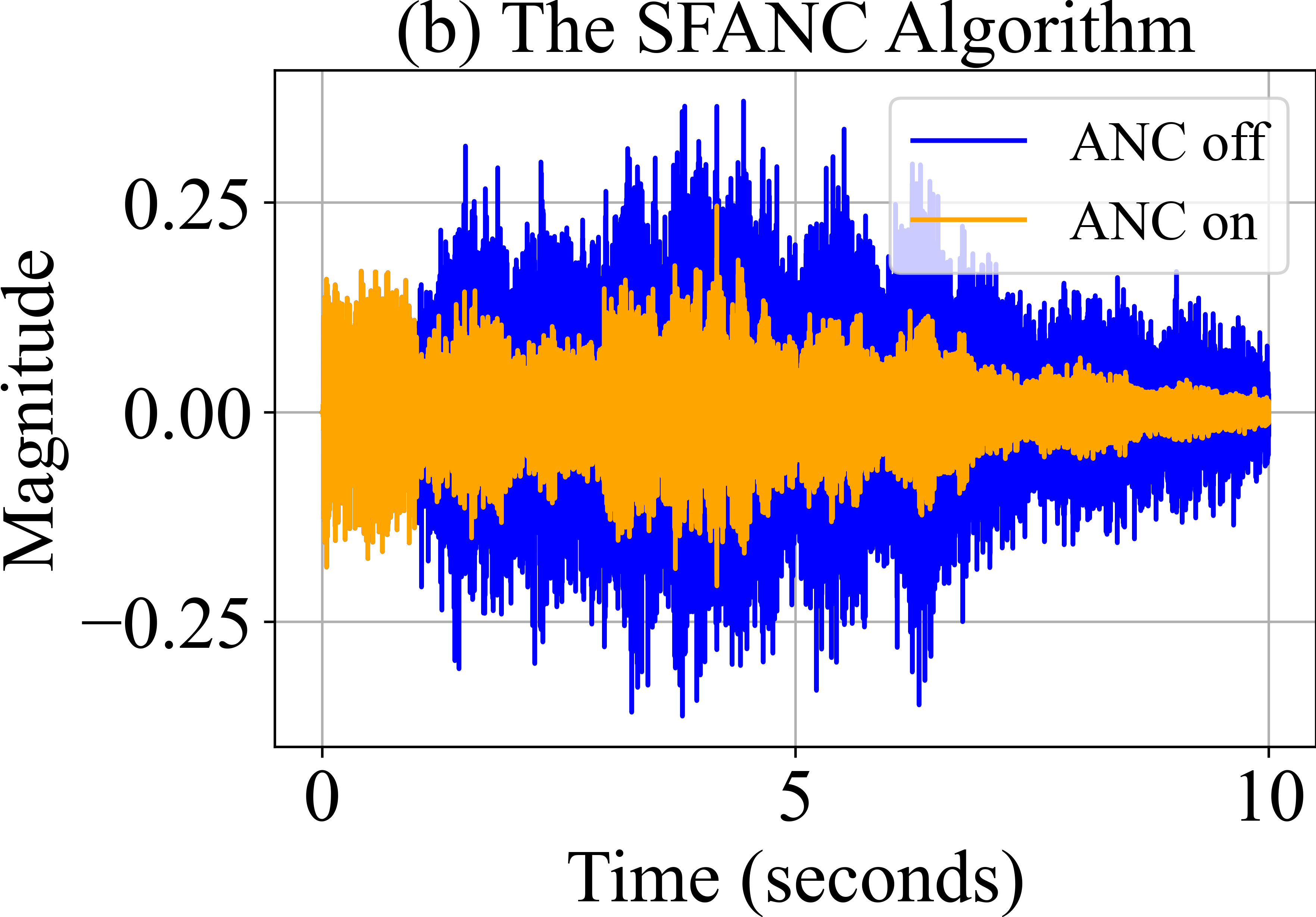}
}\vspace*{-0.3cm}
\subfigure{
\includegraphics[width=0.455\linewidth]{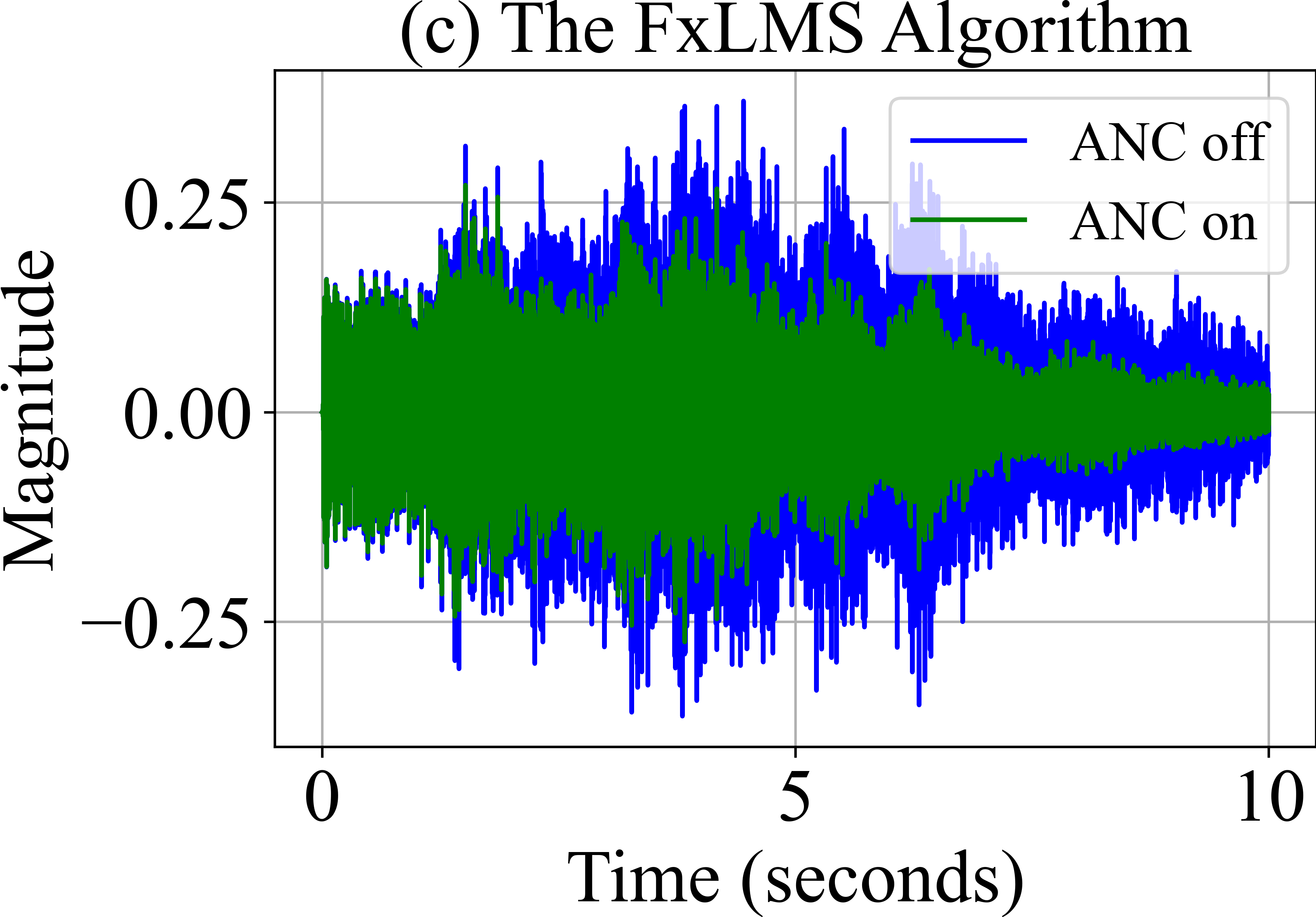}
}
\subfigure{
\includegraphics[width=0.44\linewidth]{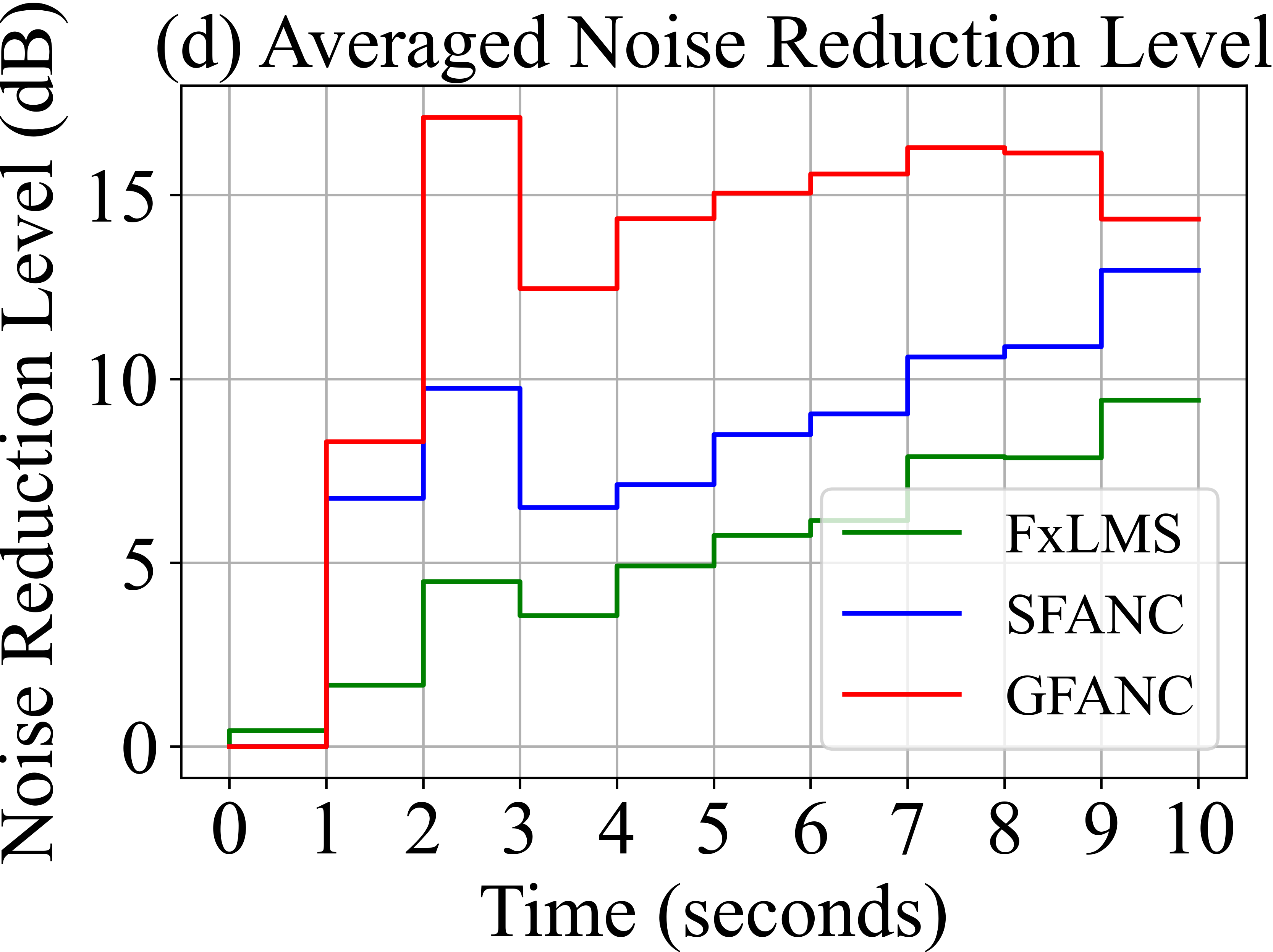}
}\vspace*{-0.3cm}
\caption{(a)-(c): Attenuated noise signals by different ANC algorithms, (d): Averaged noise reduction level on the aircraft noise ($50$Hz-$8,000$Hz) during every 1 second.}\vspace*{-0.4cm}
\label{Fig 4}
\end{figure}

\begin{figure}[!t]
\centering
\subfigure{
\includegraphics[width=0.455\linewidth]{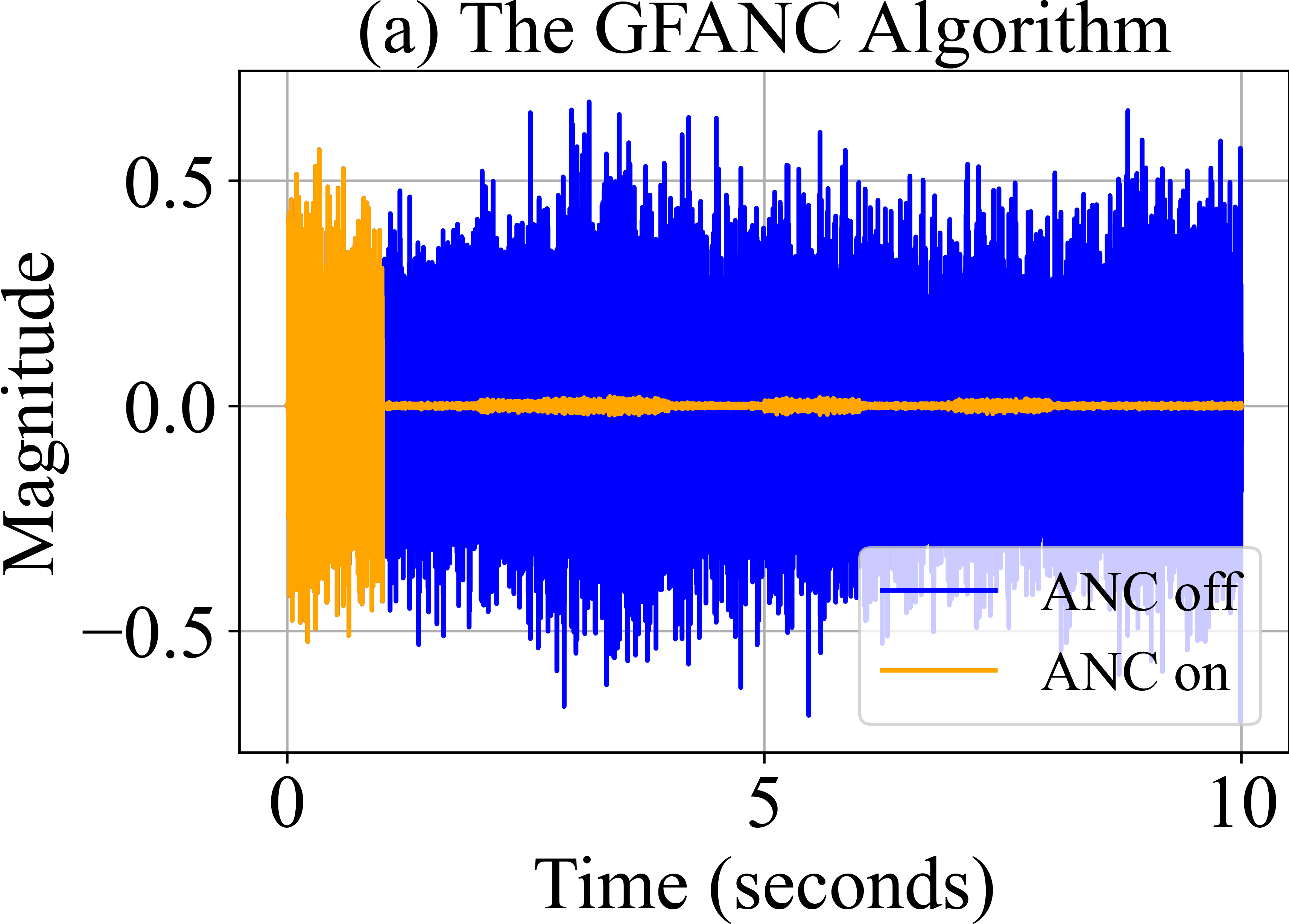}
}
\subfigure{
\includegraphics[width=0.455\linewidth]{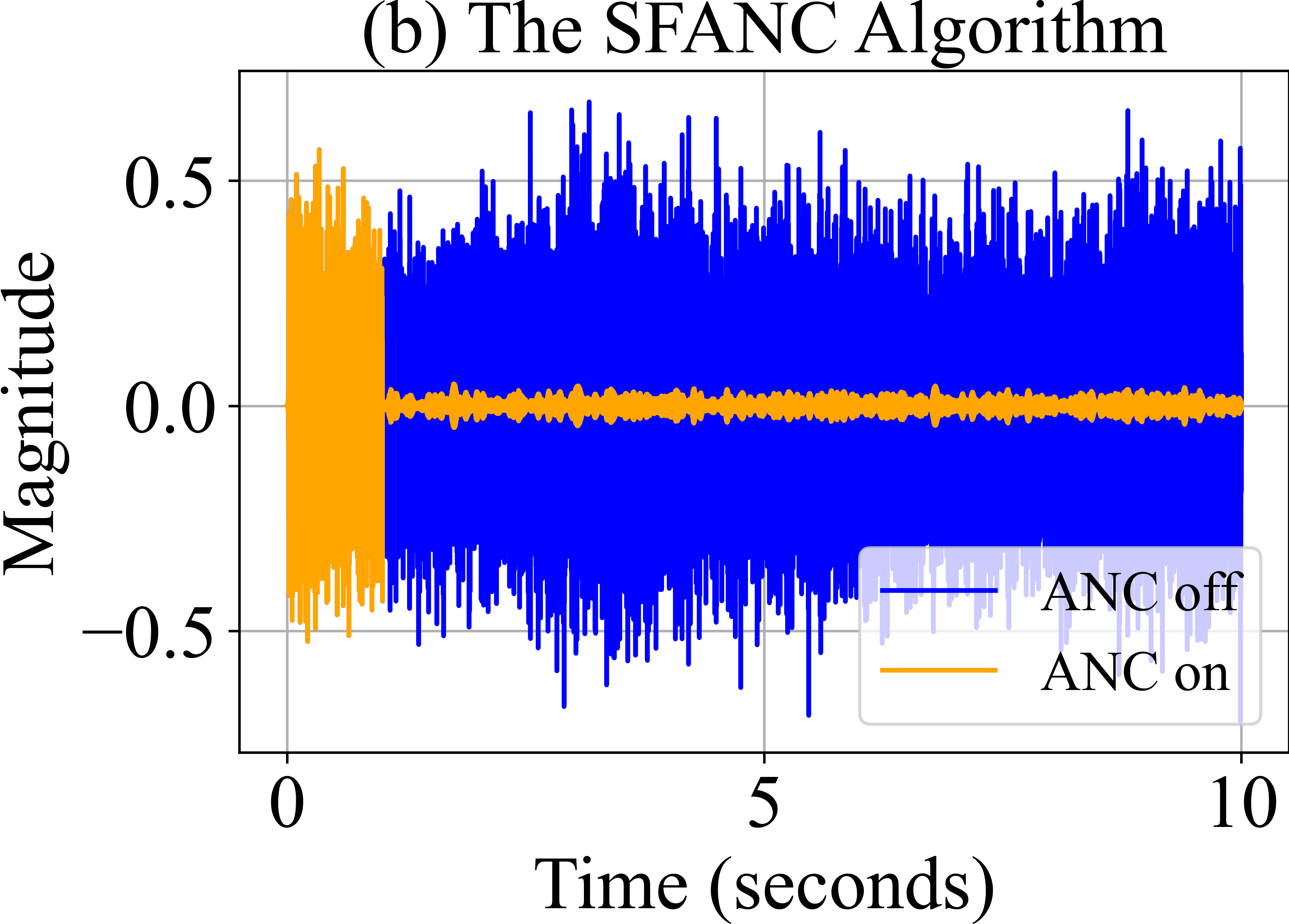}
}\vspace*{-0.3cm}
\subfigure{
\includegraphics[width=0.455\linewidth]{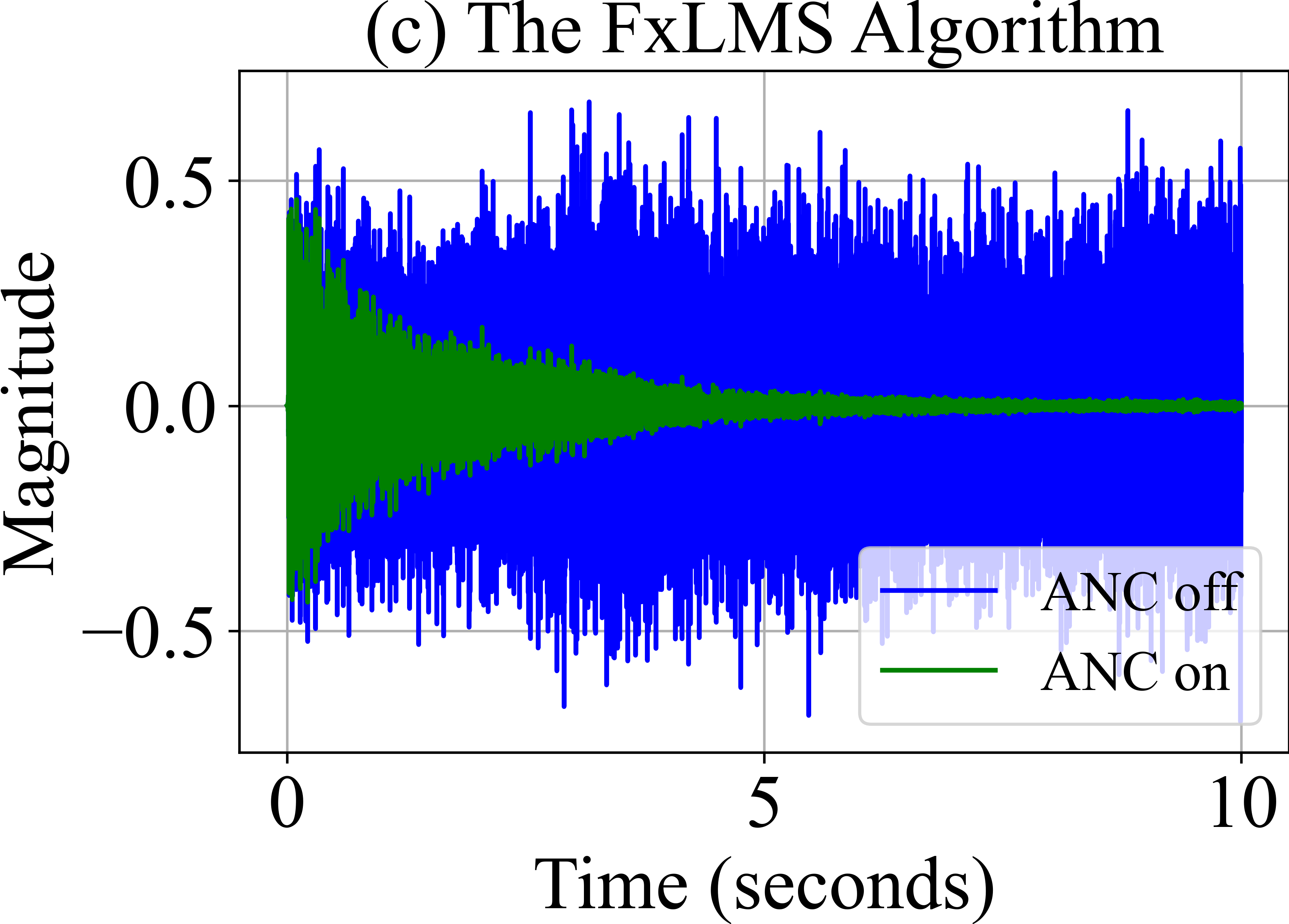}
}
\subfigure{
\includegraphics[width=0.44\linewidth]{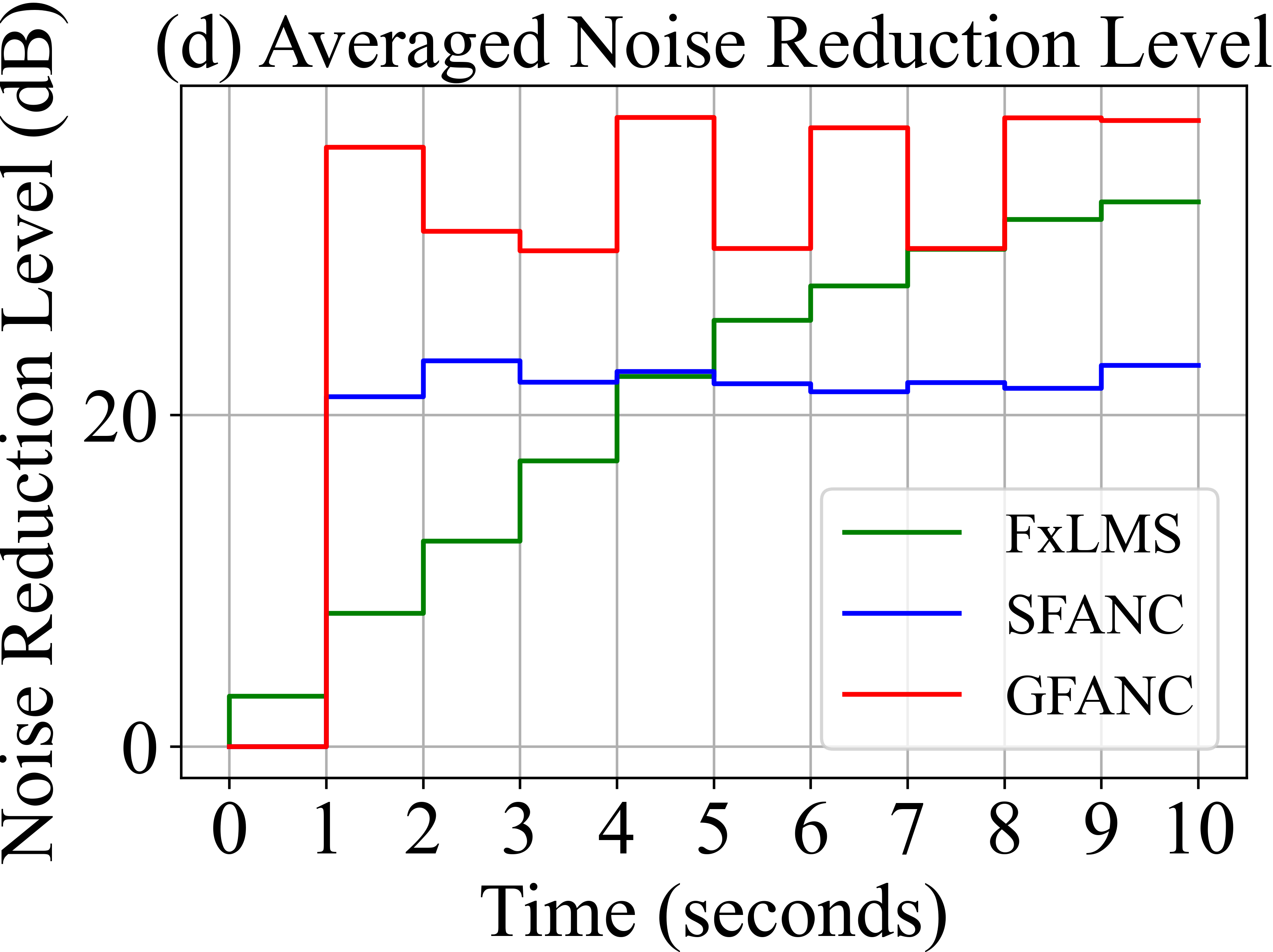}
}\vspace*{-0.3cm}
\caption{(a)-(c): Attenuated noise signals by different ANC algorithms, (d): Averaged noise reduction level on the traffic noise ($40$Hz-$1,200$Hz) during every 1 second.}\vspace*{-0.5cm}
\label{Fig 5}
\end{figure}

\subsection{Real-world Noise Cancellation}\vspace*{-0.1cm}
The GFANC algorithm is compared to SFANC \cite{11} and FxLMS in controlling real-world noises, which do not belong to the training dataset. The step size of FxLMS algorithm is $0.0001$. In the first simulation in Fig.~\ref{Fig 4}, the ANC system utilized different algorithms to cancel the aircraft noise. It is observed that the proposed GFANC algorithm outperforms SFANC and FxLMS algorithms in terms of response speed and noise reduction level. Specifically, during $2$s-$3$s, the average noise reduction level achieved by GFANC is approximately 7dB and 12dB greater than that of SFANC and FxLMS, respectively. This suggests that the GFANC algorithm is able to track and respond rapidly to varying noise. However, GFANC and SFANC cannot handle the first-second noise because they update the control filter coefficients for the next second based on the first-second noise.

Also, these ANC methods are used to attenuate the low-frequency traffic noise, as presented in Fig.~\ref{Fig 5}. Though the SFANC method responds to the noise faster than the FxLMS method in the first $4$s, the FxLMS algorithm performs better after convergence \cite{26}. In comparison, the proposed GFANC method always performs best during the noise reduction process. The results indicate that the control filter generated by the GFANC method is more suitable for the noise than that provided by the SFANC method. Furthermore, our previous research~\cite{11} demonstrated that combining fixed-filter and adaptive ANC algorithms can achieve rapid convergence and satisfactory noise reduction level.

\begin{figure}[tp]
\centering
\subfigure{
\includegraphics[width=0.474\linewidth]{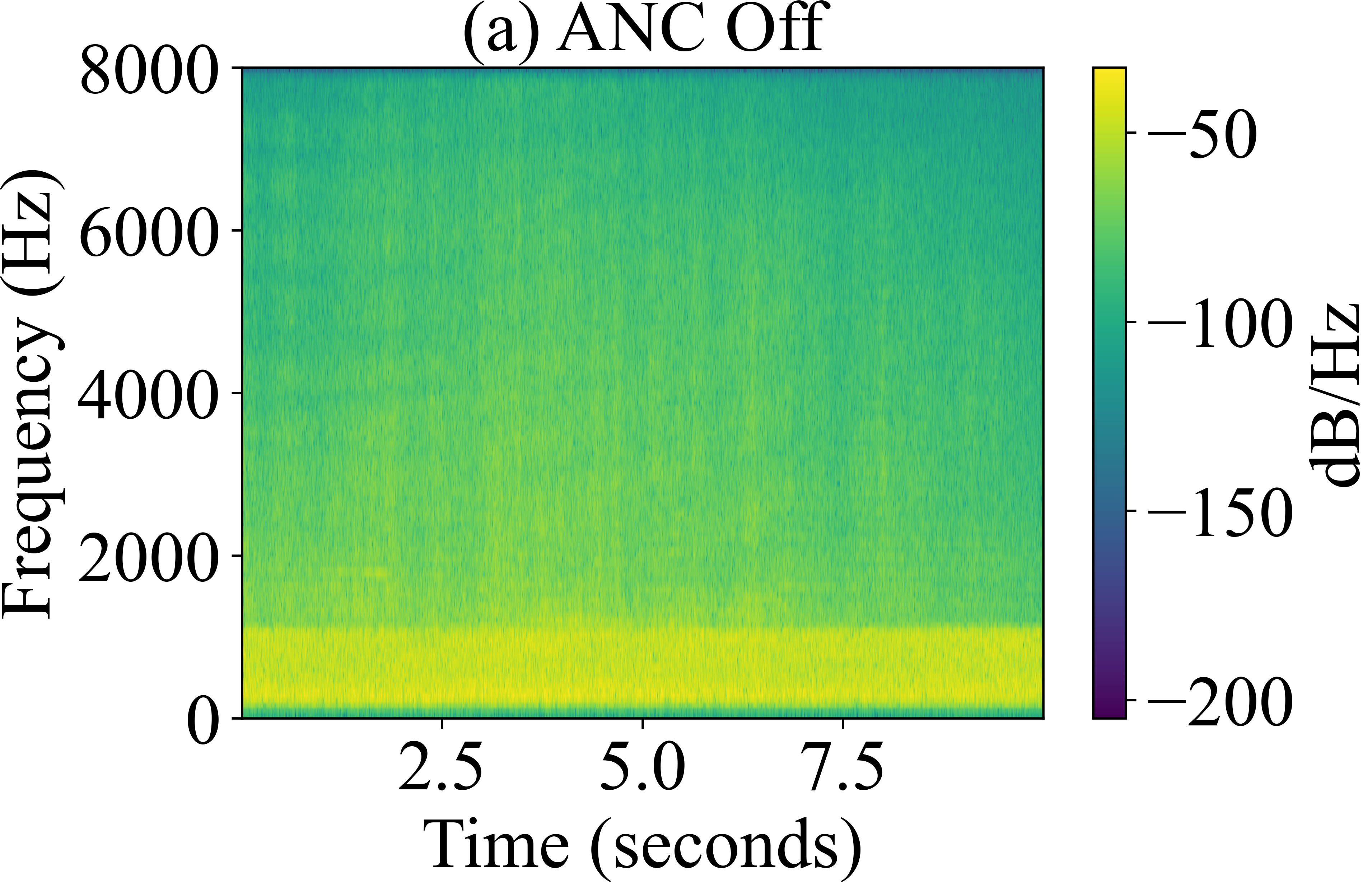}
}
\subfigure{
\includegraphics[width=0.474\linewidth]{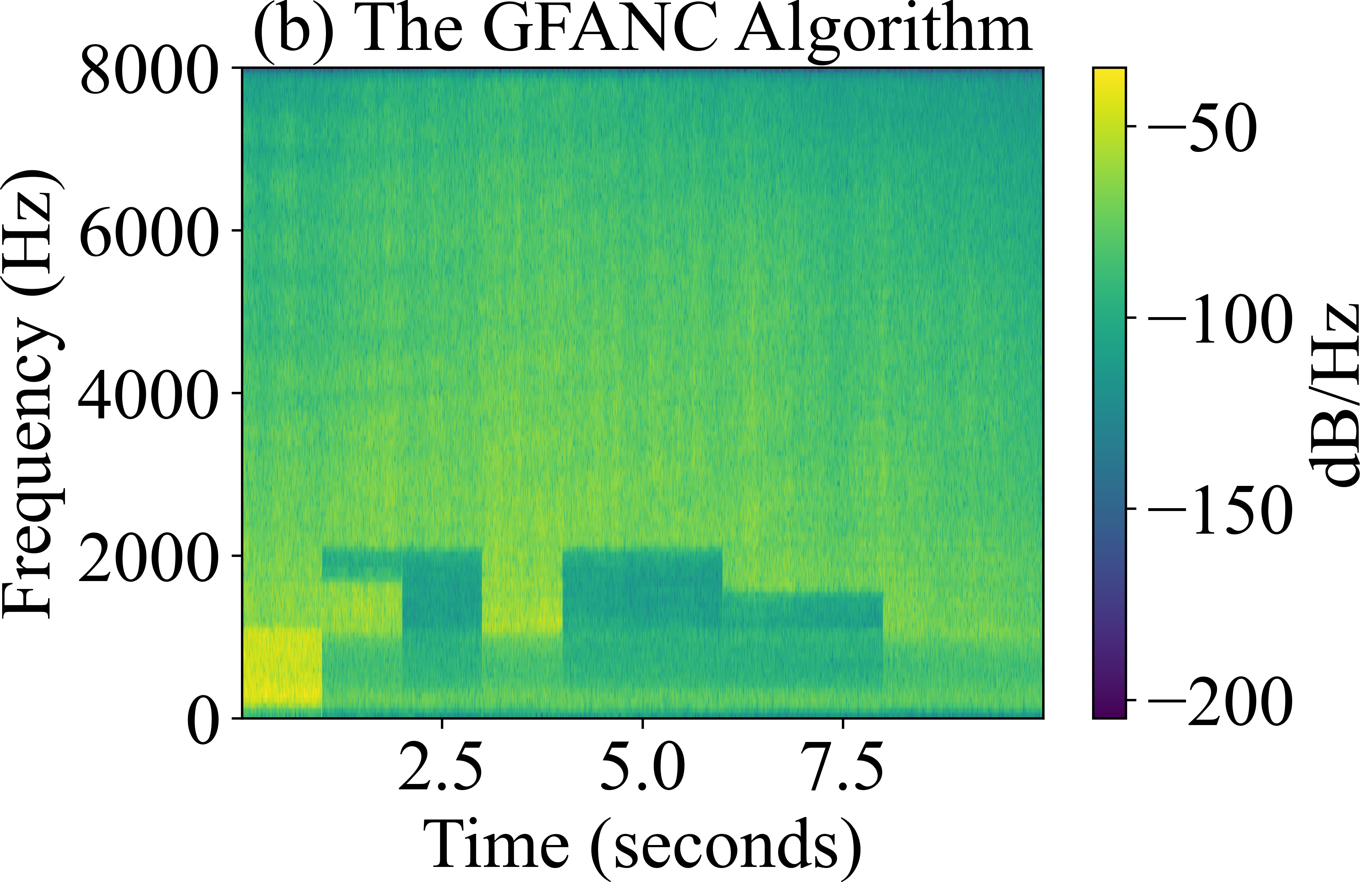}
}\vspace*{-0.2cm}
\subfigure{
\includegraphics[width=0.474\linewidth]{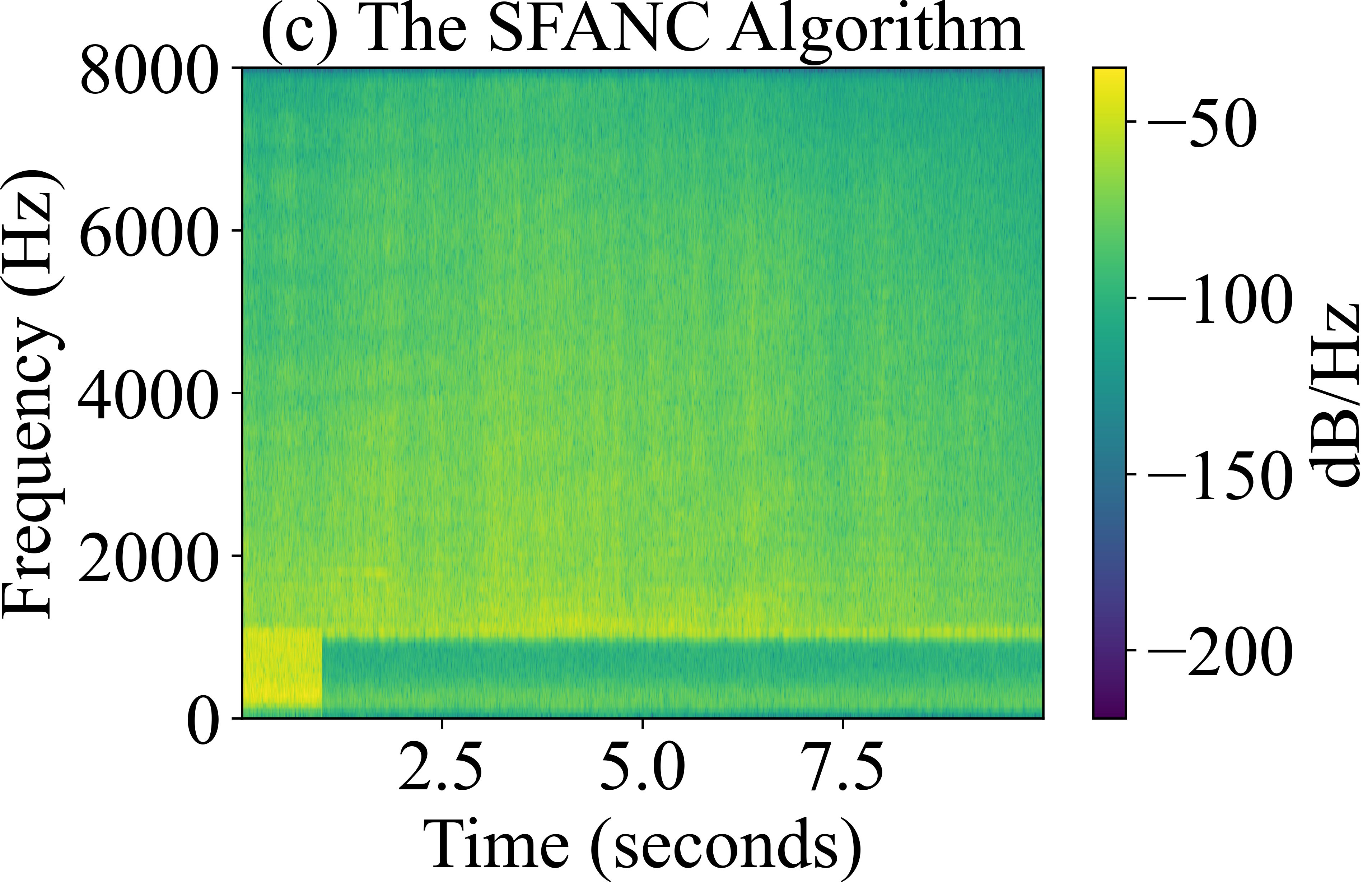}
}
\subfigure{
\includegraphics[width=0.474\linewidth]{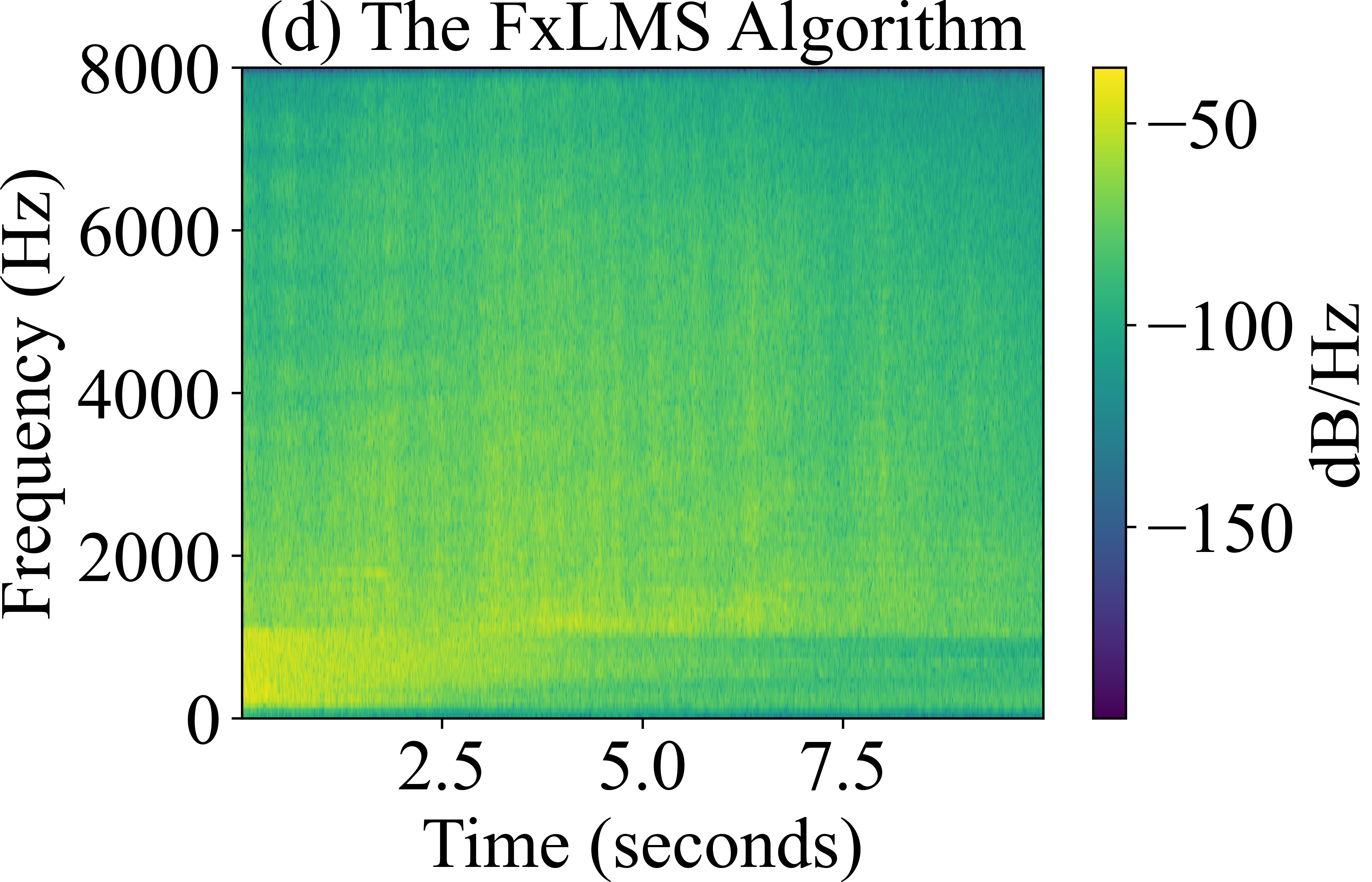}
}\vspace*{-0.3cm}
\caption{The spectrogram of noise signal obtained (a) without ANC algorithm, (b)-(d) with different ANC algorithms, on the mixed noise of aircraft and traffic noise.}\vspace*{-0.5cm}
\label{Fig 6}
\end{figure}

Furthermore, we added the aircraft and traffic noises together in time domain to get a mixed noise. The spectrograms of the mixed noise attenuated by different ANC algorithms are depicted in Fig.~\ref{Fig 6}. The spectrogram represents the power spectral density of the noise. Noticeably, most components of the mixed noise can be effectively attenuated by the GFANC algorithm. However, SFANC is less effective at removing the noise components above $1,000$Hz. The reason is that SFANC can only choose control filters from a limited number of candidates, whereas GFANC is capable of generating more appropriate control filters. Also, it is found that the FxLMS algorithm responds much slower on the mixed noise than GFANC and SFANC.\vspace*{-0.45cm}

\section{Conclusion}\vspace*{-0.25cm}
This paper proposed a novel GFANC method that generates different control filters given different primary noises. A lightweight 1D CNN operated on a co-processor extracts the noise features to form a binary weight vector. Using the weight vector, a new control filter is generated by combining sub control filters obtained from the decomposition of a pre-trained broadband control filter. Owing to the flexibility of generating control filters and avoiding optimization progress, the GFANC achieves better response speed and noise reduction level than SFANC and FxLMS. Additionally, the training dataset is automatically labelled via the adaptive mechanism instead of manual labelling. Also, only one pre-trained control filter is required, significantly reducing the effort for pre-training control filters in real-world applications.

\newpage
\bibliographystyle{IEEEbib}
\small
\bibliography{A}
\end{document}